\documentclass[aps,preprint,a4paper,amsmath,floatfix]{revtex4}
\usepackage{graphicx}

\begin{document}

\title{Phase behavior of mixtures of hard ellipses: 
A scaled particle density functional 
study}

\author{Y. Mart\'{\i}nez-Rat\'on} 
\address{Grupo Interdisciplinar de Sistemas Complejos (GISC),
Departamento de Matem\'{a}ticas, Escuela Polit\'{e}cnica Superior, Universidad Carlos III de Madrid,
Avenida de la Universidad 30, E--28911, Legan\'{e}s, Madrid, Spain}

\begin{abstract}
We present a scaled particle density functional
study of two-dimensional binary mixtures of 
hard convex particles with one or both species being ellipses. 
In particular, we divide our study into two 
parts. The first part is devoted to the calculation 
of phase diagrams of mixtures with the same elliptical shapes,  
but with 
(i) different aspect ratios and equal particle areas, (ii) equal aspect ratios 
and different particle areas and (iii) with the former and the later being 
different. In the second study we obtain the phase diagrams corresponding 
to crossed-mixtures of particles with species having elliptical 
and rectangular shapes. We compare the phase diagram topologies  
and explain their main features from the entropic nature 
of particle interactions directly related to 
the anisotropies, 
areas, and shapes of species. 
The results obtained can be corroborated by experiments on granular rods where 
the entropic forces are very important in the stabilization of 
liquid-crystal textures at the stationary states.  
\end{abstract}

\maketitle

\section{Introduction}
The demixing transition in mixtures of hard convex three-dimensional particles 
have received a considerably attention from the theoretical point 
of view 
\cite{Roij,Wensink,Dubois,Perera,Varga,Galindo,M-R1,Schmidt,Schmidt1,Roij2,Roij3,Dijkstra,Varga2,Purdy}. 
This is due to the possibility of formulating relatively simple 
theoretical models for mixtures of anisotropic particles which  
are not allowed to overlap. These models are based on the simple Onsager 
second-virial approximation \cite{Onsager} or its rescaled version, the Parsons 
approach \cite{Parsons}. Also, the scaled particle theory (SPT), proposed 
firstly for hard spheres \cite{Reiss} and later extended to other 
particles geometries \cite{Cotter,Cotter1,Cotter2,Lasher,Barboy}, has been a fruitful theoretical tool in 
the study of these kind of mixtures. All these models predict  
that a binary mixture of asymmetric enough particles  
can demix into two different phases, each one rich in different species. 
The nature of uniform demixed phases can be isotropic (I), a
fluid with randomly oriented particle axis, or nematic (N) in which
the particles' axes 
are oriented in average along the nematic director. All possible demixing 
transitions, i.e. I-I, N-N and I-N were found from the numerical 
implementation of these models. 

In two dimensions the theoretical works devoted to the study 
of the demixing 
transitions are more scarce, because   
strictly two-dimensional 
fluids occur less often in nature compared to the three-dimensional fluids. Examples of the  
former constitute the quasi-two-dimensional films obtained from 
preferential chemical adsorption of certain species at interfaces.  
The adsorption of polystyrene spheres  
\cite{Pieranski} 
or ellipsoidal latex particles \cite{Basavaraj} into a two-dimensional 
interface has allowed the experimental study of  
freezing and percolation/jamming transitions in two dimensions respectively.
Also, the time-dependent behavior of anisotropic 
colloids near interfaces was recently studied through 
a theoretical modelling of the Langevin dynamics of adsorbed particles 
at a flat interface \cite{Graaf}. For a recent review on the behavior of 
colloidal particles at liquid interfaces see Ref. \cite{Binks}.

However, the most paradigmatic example constitutes the 
collection of granular anisotropic particles confined between two parallel 
plates with the distance between them less than the particle widths.  
A growing number of experimental works 
conducted on granular  
fluids subjected to different vibrational modes show
spatial pattern formations, demixing scenarios of 
granular species and collective motions \cite{Aronson}. 
Those systems in which 
granular rods are vertically shacked  
have shown 
that the entropic nature of interactions between  
particles is very important in the explanation of  
phenomenology found. In particular, the 
presence of defects of the nematic director due to the difference between 
particle orientations near and far from 
the walls was found \cite{Galanis}.  Recent experiments on granular rods 
reported the presence of  
certain liquid-crystal 
textures with N, tetratic (T), and smectic (Sm) symmetries 
as the stationary states of these systems \cite{Narayan}.   
The presence of strong dissipation due to the 
inelastic interactions between grains is an important 
ingredient to take into account in any theoretical model 
of a shaken granular systems. However, the presence  
of demixed stationary states of a 
granular mixture could be described 
by models based of density functional theory. For example 
the ranges of molar and packing fractions for which 
the demixing is observed could be approximately given by 
these models.  

The SPT shows that a two-dimensional mixture of hard convex bodies never 
demixes into two I phases \cite{Talbot}. 
However, recently we have shown that the 
same theory can predict I-N and N-N demixing 
\cite{M-R2,Heras}. 
The only difference with respect to three dimensions is 
the second order character of 
the I-N transition at pressures bellow the tricritical point 
\cite{M-R2,Heras}. The main purpose of the present study is to implement 
the SPT for (i) a mixture of hard ellipses (HEs) with different aspect 
ratios and/or particle areas and (ii) for crossed-mixtures composed 
of HEs and hard rectangles (HR). The 
most convex particle with a fixed aspect ratio that can be imagined 
in two dimensions is just an 
ellipse. This property in turn might be reflected in the phase 
behavior of mixtures composed of HEs. We will show that this is indeed 
the case by studying the phase diagram topology as a function of 
particle aspect ratio and particle area. The one-component fluid 
of HEs has been studied via Monte Carlo simulations.   
This fluid exhibits, for high enough aspect ratios, an I-N transition  
occurring via the disclination unbinding mechanism, while for 
higher packing fractions a first order fluid-solid transition takes place \cite{Cuesta}. 
It has been shown that the equation of state for the one-component fluid 
of HEs following the SPT compares reasonable well with 
the simulation results \cite{Schlacken}.

To implement the present study we firstly find an analytical expression 
for the excluded area between two HEs with different axes 
$\{{\sf a}_i,{\sf b}_i\}$ ($i=1,2$). Zheng and Palffy-Muhoray \cite{Zheng} 
developed a method based 
on the calculation of the distance of closest approach
between an ellipse and 
a circle. Thus, after rescaling the circle they obtained 
the distance of closest approach between two ellipses. Finally, the 
excluded area can be obtained by the angular integration of the distance  
of closest approach. However, this integral 
should be evaluated numerically. We will follow another procedure,  
in which the excluded area between 
two different ellipses 
can be calculated from the parametrization of the distance of closest 
approach as a function of the angle that forms the point of contact 
between two ellipses with a fixed reference axis. The resulting  
excluded area thus depends on the complete elliptic integral 
with its parameter being a function of the ellipse eccentricities 
and on the relative angle between particles. Further, we also calculate 
the excluded area between an ellipse and a rectangle. The details of 
both calculations are relegated to the appendices.  

This paper is organized as follows. In Sec. \ref{SPT} we present the 
theoretical model used for the present study: the SPT. In Sec. \ref{bifurca} we 
analytically derive the spinodal curves of the I-N second-order 
transition, i.e. 
the packing fraction as a function of 
composition for general mixtures of 
two-dimensional hard particles . Sec. \ref{NN} is devoted to 
the analytical derivation of the N-N 
spinodal curve with the constraint of parallel particle alignment. 
In Sec. \ref{results} we present the phase diagrams of mixtures of HEs 
(Sec. \ref{HE}) and 
HEs and HRs (Sec. \ref{cross}) 
obtained from the numerical minimization of the free-energy functional. 
Finally, in Sec. \ref{conclusions} the main conclusions are drawn. The 
appendices contain the 
details of the derivation of the excluded area between two HEs 
(Sec. \ref{A_HE}) and 
HEs and HRs (Sec. \ref{A_HE_HR}) and different mathematical aspects of the 
functional minimization procedure (Sec. \ref{minimiza}).

\section{Scaled particle theory}
\label{SPT}
The central quantity of any theoretical model which accounts for 
two-dimensional hard particle interactions 
is the pair excluded area to particle $\nu$ 
due to the presence of particle $\mu$ with the constraint 
of a fixed angle $\phi$ between their main axis. Because the hard core 
nature of interactions, both particles can not 
overlap and thus the excluded area $A_{\mu\nu}(\phi)$ defines a body with 
a geometry directly related with those of particles.   
The present paper is devoted to the study of phase diagrams of mixtures  
composed of HEs with different characteristic lengths [different 
semi-axis $\left({\sf a}_{\mu},{\sf b}_{\mu}\right)$] and a crossed--mixture 
of HEs and HRs. In Appendices \ref{A_HE} and \ref{A_HE_HR} we obtain 
the analytical formulas 
for the excluded areas of these mixtures.  
Once the excluded area is given we can construct a scaled 
particle theory (SPT) for a general mixture \cite{Reiss}.  
The excess free energy per particle 
following 
the SPT can be calculated to be
\begin{eqnarray}
\varphi_{\rm{ex}}=-\ln(1-\eta)+\frac{\rho}{1-\eta}\sum_{\mu\nu}x_{\mu}
x_{\nu}\langle\langle A^{(0)}_{\mu\nu}\rangle\rangle,
\label{gg}
\end{eqnarray}  
where $\rho$ is the total number density and $\eta=\rho\sum_ix_i a_i$ is the 
total packing fraction with $x_i$ the molar fraction of species $i$ and 
$a_i$ their particle areas. 
We have defined the function 
$A^{(0)}_{\mu\nu}(\phi)=\left(A_{\mu\nu}(\phi)-a_{\mu}-a_{\nu}\right)/2$, 
and used the short-hand notation 
\begin{eqnarray}
\langle\langle A^{(0)}_{\mu\nu}\rangle\rangle=\int_0^{\pi}
d\phi_1\,h_{\mu}(\phi)\int_0^{\pi}d\phi_2\,h_{\nu}(\phi)
A^{(0)}_{\mu\nu}(\phi_1-\phi_2),
\label{sd}
\end{eqnarray}
to define the double angular average of  
$A^{(0)}_{\mu\nu}(\phi)$ with respect to the orientational 
distribution functions $h_{\mu}(\phi)$. Note that the second 
order density expansion of Equation (\ref{gg}) give us the usual 
second virial approximation 
\begin{eqnarray}
\varphi_{\rm{ex}}^{(\rm{b}_2)}=\frac{\rho}{2}\sum_{\mu\nu}x_{\mu}x_{\nu}
\langle \langle A_{\mu\nu}\rangle\rangle.
\end{eqnarray}
The fact that the coefficients $\langle\langle A^{(0)}_{\mu\nu}\rangle\rangle$, 
instead of the excluded areas, enter into the SPT free-energy   
is due to the form of the derived SPT: through a Taylor expansion of 
the excess chemical potential 
corresponding to the scaled particle, evaluated at low densities. 
The upper 
limit of integration in Equation (\ref{sd}) is $\pi$ due to the head-tail symmetry of 
particles, i.e. $h_{\mu}(\phi)=h_{\mu}(\pi-\phi)$.  These 
functions measure the degree of orientation of the particles 
along the fixed nematic director. In the 
present paper we have selected a variational family 
for these functions as
\begin{eqnarray}
h_{\mu}(\phi)=\frac{\exp\left(\lambda_{\mu}\cos 2\phi\right)}
{\pi I_0(\lambda_{\mu})},
\label{ins}
\end{eqnarray}
with $I_n(x)$ the n-th order modified Bessel function.  
The perfect nematic alignment is reached when $\lambda_{\mu}\to\infty$, 
which gives us $h_{\mu}(\phi)=\delta(\phi)$ (the 
Dirac-delta function), 
while the isotropic fluid is obtained in the limit $\lambda_{\mu}\to 0$ 
for which $h_{\mu}(\phi)=\pi^{-1}$.  
Note that this 
parametrisation excludes the possibility of study of the tetratic phase with 
fourfold angular symmetry $h_{\mu}(\phi+\pi/2)=h_{\mu}(\phi)$.  

The ideal part of the free-energy per particle can be computed as 
\begin{eqnarray}
\varphi_{\rm{id}}&=&\ln \eta -1+\sum_{\mu}x_{\mu}\left\{\ln x_{\mu}+
\int_0^{\pi}d\phi h_{\mu}(\phi)\ln\left[\pi h_{\mu}(\phi)\right]\right\}, 
\label{lp1}\\
&=&\ln \eta -1+\sum_{\mu}x_{\mu}\left\{\ln x_{\mu}+
\lambda_{\mu}\frac{I_1(\lambda_{\mu})}{I_0(\lambda_{\mu})}-
\ln I_0(\lambda_{\mu})\right\}.
\label{lp2}
\end{eqnarray}
where we have inserted (\ref{ins}) into (\ref{lp1}) to obtain 
(\ref{lp2}).

The Gibbs free energy per particle $g=\varphi+P/\rho$ (with 
$\varphi=\varphi_{\rm{id}}+\varphi_{\rm{exc}}$ the total free-energy 
per particle) 
calculated 
at a constant fluid pressure $P$ 
should be minimized with 
respect to the parameters  $\lambda_{\mu}$ ($\mu=1,2$) 
to obtain their equilibrium values.
The expression for the pressure following the
SPT is
\begin{eqnarray}
\beta P&=&\frac{\rho}{1-\eta}+\frac{\rho^2}{(1-\eta)^2}
\sum_{\mu\nu}x_{\mu}x_{\nu}\langle\langle A^{(0)}_{\mu\nu}\rangle\rangle.
\label{can}
\end{eqnarray}
In Appendix \ref{minimiza} we write explicit expressions for the 
equations to be solved numerically to obtain the equilibrium values of 
$\lambda_{\mu}$.   
The common tangent construction 
of the function $g(x)$, with $x=x_1$, allows us to calculate 
the coexisting values 
for $x$ (and thus of $\eta$ once the fluid pressure is fixed) at
the demixing transition.  

\section{I-N bifurcation analysis}
\label{bifurca}
In this section we obtain an analytic expression for the 
I-N spinodal of a general mixture. 
When $\lambda_{\mu}\ll 1$ we can approximate the orientational 
distribution functions $h_{\mu}(\phi)$ up to first order in  
$\lambda_{\mu}$ by
\begin{eqnarray}
h_{\mu}(\phi)=\frac{1}{\pi}\left(1+\lambda_{\mu}\cos 2\phi\right).
\end{eqnarray}
Inserting this expression in the free-energy per particle 
$\varphi=\varphi_{\rm{id}}+\varphi_{\rm{exc}}$ we obtain, 
up to second order in $\lambda_{\mu}$
\begin{eqnarray}
\varphi= \frac{1}{4}\sum_{\mu}x_{\mu}\lambda_{\mu}^2+
\frac{y}{2}\sum_{\mu\nu}
x_{\mu}x_{\nu}\left[\alpha^{(0)}_{\mu\nu}+
\alpha^{(1)}_{\mu\nu}\lambda_{\mu}
\lambda_{\nu}\right],
\label{tt}
\end{eqnarray} 
with $y=\rho/(1-\eta)$ and where we have defined the coefficients
\begin{eqnarray}
\alpha^{(i)}_{\mu\nu}=\frac{1}{\pi}\int_0^{\pi}d\phi
\cos(2i\phi)A_{\mu\nu}^{(0)}(\phi),\quad i=0,1.
\end{eqnarray}
The minimization of (\ref{tt}) with respect to $\lambda_{\mu}$ 
give us
\begin{eqnarray}
\frac{\partial \varphi}{\partial\lambda_{\mu}}=
\frac{x_{\mu}}{2}\left[\lambda_{\mu}+2y\sum_{\nu}\alpha_{\mu\nu}^{(1)}
x_{\nu}\lambda_{\nu}\right]=0.
\label{ss}
\end{eqnarray}
The term of Eq. (\ref{ss}) enclosed by square brackets 
can be put in the following matrix form  
\begin{eqnarray}
H \boldsymbol{\lambda}\equiv \left(I+2y U\right)\boldsymbol{\lambda}={\bf 0},
\label{mm}
\end{eqnarray}
where
$\boldsymbol{\lambda}\equiv 
(\lambda_1,\lambda_2)^T\neq {\bf 0}$, $I$ is the $2\times 2$ identity matrix 
while $U$ is the matrix with elements $\alpha_{\mu\nu}^{(1)}x_{\nu}$.  
Equation (\ref{mm}) has a nontrivial solution only if 
$\rm{det}\left(H\right)=0$, which gives us the following result:
\begin{eqnarray}
1+2y\tau
+(2y)^2x_1x_2\left\{\alpha_{11}^{(1)}\alpha_{22}^{(1)}
-\left[\alpha_{12}^{(1)}\right]^2\right\}=0,
\label{ll}
\end{eqnarray}
where we have defined the coefficient 
\begin{eqnarray}
\tau=\sum_{\mu}x_{\mu}\alpha_{\mu\mu}^{(1)}.
\end{eqnarray}
Taking into account that the expression enclosed by brackets in Equation 
(\ref{ll}) is equal to zero for any mixture of convex bodies, we find 
that the packing fraction at the I-N transition can be calculated as 
\begin{eqnarray}
\eta^*&=& \frac{y^*\langle a\rangle}{1+y^*\langle a\rangle}, 
\quad y^*=-\frac{1}{2\tau},
\end{eqnarray}
where we have used the shorthand notation $\langle a\rangle 
=\sum_{\mu}x_{\mu}a_{\mu}$.
Following the SPT the pressure of the isotropic fluid at the 
bifurcation point can be 
calculated as
\begin{eqnarray}
\beta P^*&=&y^*\left\{1+\frac{1}{4\pi}y^*\left[\sum_{\mu}x_{\mu}{\cal L}_{\mu}
\right]^2\right\}, \label{p1}
\end{eqnarray}
where ${\cal L}_{\mu}$ is the perimeter of species $\mu$.

\section{N-N demixing}
\label{NN}
\begin{figure}
\begin{center}
\mbox{\includegraphics*[width=3.5in, angle=0]{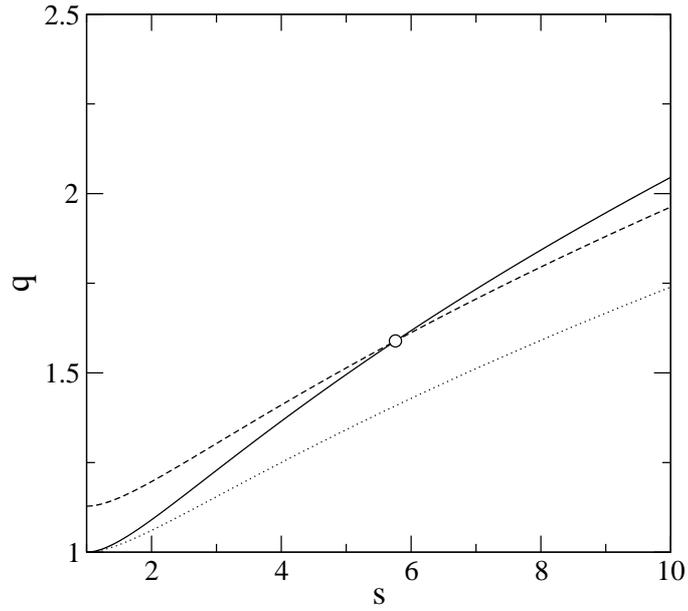}}
\end{center}
\caption{The coefficient $q$ as a function of $s=\kappa_2/\kappa_1$ for 
a mixture of HEs (solid curve), for the crossed-mixture of 
HEs and HRs (dashed curve), and for a mixture 
of HRs (dotted curve). The open circle shows the 
cross-over point.}  
\label{fig0}
\end{figure}

\begin{figure}
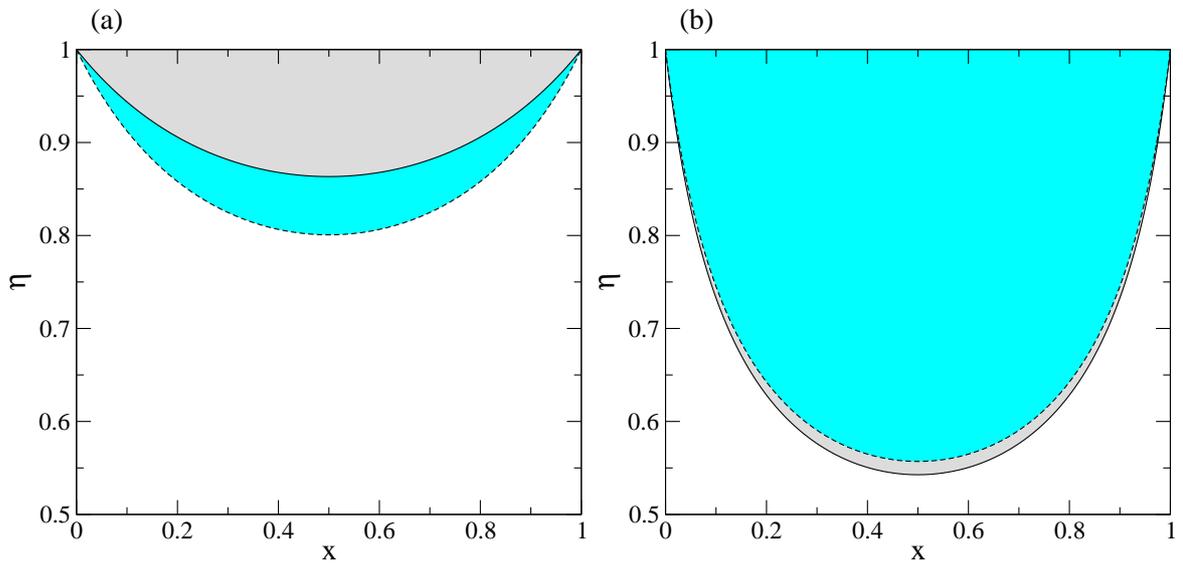

\begin{center}
\mbox{\includegraphics*[width=3.in, angle=0]{Fig2a.eps}}
\mbox{\includegraphics*[width=3.in, angle=0]{Fig2b.eps}}
\end{center}
\caption{Demixing spinodals of mixtures of HEs (solid line) 
and of the crossed-mixture of HEs and HRs (dashed line) for 
$s=2.5$ (a) and $s=8$ (b).} 
\label{demix}
\end{figure}

In this section we obtain analytically the criterion for the existence of 
N-N demixing and its spinodal curve with the constraint of 
parallel particle alignment. This constraint is necessary to obtain 
an explicit analytic function for this spinodal, while the freely 
rotating case can only be solved numerically.  
The spinodal instability of the binary mixture with respect to  
phase separation can be computed by requiring that 
${\cal H}(\eta,x)=0$ where we have defined  
\begin{eqnarray}
{\cal H}(\eta,x)\equiv \left[1+\rho_1\frac{\partial^2\Phi}{\partial \rho_1^2}
\right]
\left[1+\rho_2 \frac{\partial^2\Phi}{\partial \rho_2^2}\right]-
\rho_1\rho_2\left[\frac{\partial^2\Phi}{\partial\rho_1\partial\rho_2}\right]^2,
\label{gg1}
\end{eqnarray}
with $\rho_i=x_i\rho$, 
while $\Phi=\rho\varphi_{\rm{ex}}$ is the excess part of 
the free energy density.
From Equations (\ref{gg}) and (\ref{gg1}) we obtain 
\begin{eqnarray}
{\cal H}(\eta,x)=(1+y_0)^2\left\{1+2y_0-4x(1-x)y_0^2{\cal U}\right\},
\end{eqnarray}
where we have defined $y_0=\eta/(1-\eta)$, and 
${\cal U}=\left(q^2-1\right)a_1a_2/
\langle a\rangle^2$, with $q=A_{12}^{(0)}/\sqrt{a_1a_2}$. Solving 
${\cal H}(\eta,x)=0$ for $\eta$, we find the explicit expression 
\begin{eqnarray}
\eta=\frac{1}{\sqrt{1+4x(1-x){\cal U}}}.
\label{ml}
\end{eqnarray}
The packing fraction and composition at the critical point are $\eta^*=1/q$,
and $x^*=a_2/(a_1+a_2)$, respectively. 
Finally, the pressure at this point is
\begin{eqnarray}
\beta P^*a_2=q\left[\frac{1+\sqrt{a_2/a_1}}{q-1}\right]^2.
\end{eqnarray}
We conclude that demixing transition 
occurs only if ${\cal U}>0$, i.e. for 
$q> 1$. 
For mixtures of HEs we find  
$q_{\rm{ee}}=2\sqrt{s}E\left[1-s^{-2}\right]/\pi\geq 1$,
where we have defined the coefficient
\begin{eqnarray}
s=\kappa_2/\kappa_1\geq 1, 
\end{eqnarray}
with $\kappa_i$ the aspect ratio 
of particle $i$ while $E[x]$ is the complete elliptic integral of the second kind. 
The same coefficients for a mixture of HRs and for a 
crossed-mixture of HEs and HRs are   
$q_{\rm{rr}}=\left(\sqrt{s}+1/\sqrt{s}\right)/2\geq 1$
and 
$q_{\rm{er}}=\left(\sqrt{s}+1/\sqrt{s}\right)/\sqrt{\pi}\geq 2/\sqrt{\pi}>1$ 
respectively.
We see that for $s=1$ (no matter 
what the ratio $a_2/a_1$ is) no demixing occurs in mixtures of 
particles with the same shape, while the crossed-mixture demixes even 
in this case.
In Figure \ref{fig0}
we plot the coefficients $q_{\alpha\beta}$ as a function of 
$s$. 
It is interesting to note that at $s\approx 5.759$ two of the 
curves intersect (those corresponding to the mixture of HEs and 
the crossed-mixture).
To illustrate the effect of this cross-over on the phase behavior 
we plot in Figure \ref{demix} the spinodals of both types of 
mixtures for $s=2.5$ and $s=8$, which are located at different sides of 
the cross-over point.  

\section{Results}
\label{results}
All the phase diagrams we present in this section were calculated via 
the bifurcation analysis (see Section \ref{bifurca}) for the case of 
continuous phase transitions (as the second order I-N transition) 
and from the numerical minimisation of the parametrised Gibbs 
free energy per particle (see Section \ref{SPT} and Appendix \ref{minimiza}) 
for the case of I-N or N-N 
demixing transitions. The later procedure allows us to obtain the 
equilibrium orientational distribution functions $h_{\mu}(\phi)$ 
at coexistence and the coexisting values of the composition and packing fraction  
of the demixed phases at 
a given pressure. We begin the presentation of the results from the 
most simple binary mixture: a mixture of HEs. 
Note that the hard disk (HD) is a special case of ellipse with equal 
semi-axis. 

\begin{figure}
\begin{center}
\mbox{\includegraphics*[width=3.5in, angle=0]{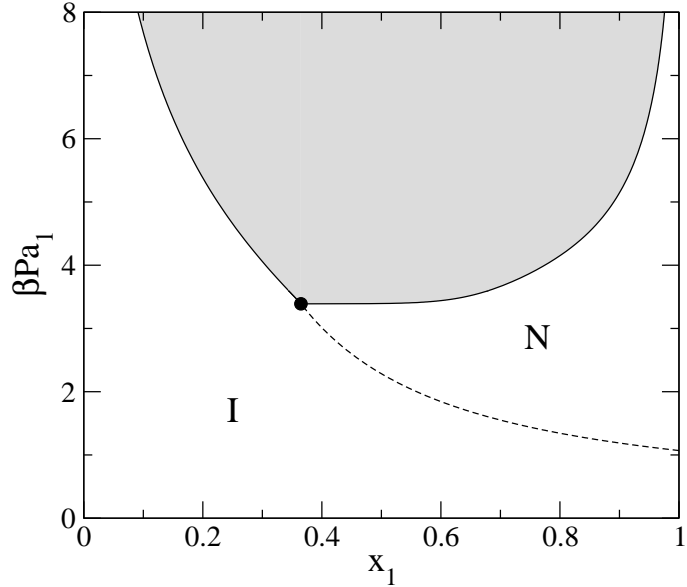}}
\end{center}
\caption{Phase diagram of the binary mixture of HEs (species 1) 
with $\kappa_1=10$ and 
HDs (species 2). Both species have the same particle areas 
$a_1=a_2$. The regions of I and N stability are correspondingly 
labeled. The I-N spinodal is plotted with dashed line while the demixing 
binodals are shown with solid lines. The grey region constitutes 
the zone of mixture instability. 
The tricritical point is shown with a filled circle.}
\label{fig1}
\end{figure}

\subsection{Mixtures of HE}
\label{HE}
The phase diagram topologies we show here  
are similar to those obtained for mixtures of HRs and 
hard discorectangles (HDRs) in recent studies \cite{M-R2,Heras}.  
These studies have shown all possible 
demixing scenarios that 
mixtures of two-dimensional hard convex particles can exhibit. However, 
we will show that the ellipse, being the geometry with the 
most convex 
anisotropy we can imagine, considerably enhance the demixing transition  
in mixtures of particles with elliptical shapes. 

The first binary mixture we study is composed of HEs 
and HDs. The former has an aspect ratio of $\kappa_1=10$ while the  
area of both species are the same ($a_1=a_2$). 
The resulting phase diagram is shown in 
Figure \ref{fig1} as a pressure--HE molar fraction ($x_1$) plot. As 
we can see from this figure, at low pressures the mixture exhibits a 
second order I-N phase transition up to 
pressure value of $\beta Pa_1\approx 3.7$ 
(the location of a tricritical point)
above which the system demixes into an I phase rich in HDs and a N phase 
rich in HEs. The demixing gap is wider as 
the pressure increases from the tricritical point. We should point out 
that the second order I-N transition obtained from density functional 
calculations replace the Kosterlitz-Thouless I-N transition obtained by 
simulations and theoretical models that properly account for the nematic 
director fluctuations. The SPT theory does not includes these fluctuations, 
thus predicting a continuous increase of the long range nematic ordering.    

\begin{figure}
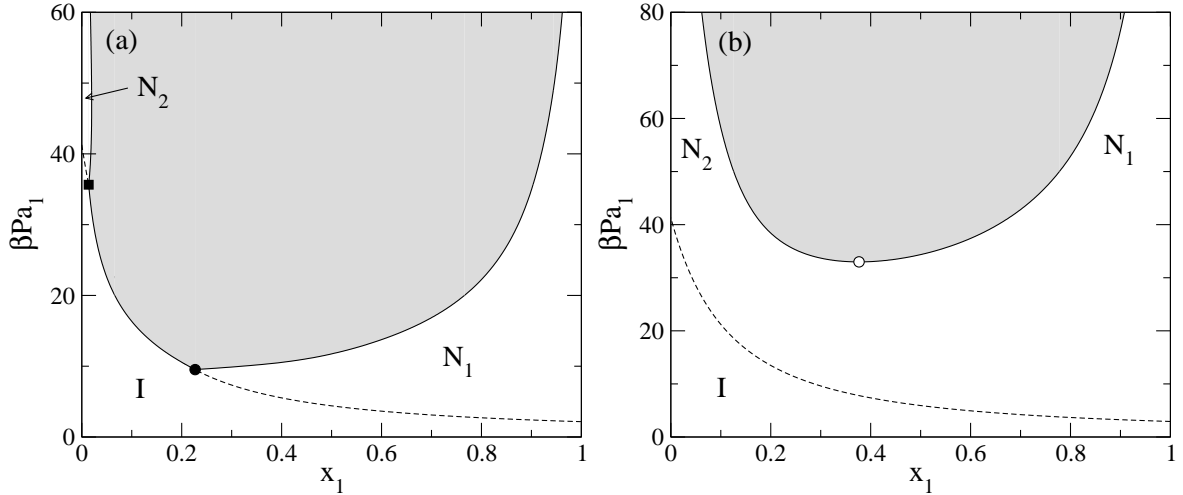

\begin{center}
\mbox{\includegraphics*[width=3.in, angle=0]{Fig4a.eps}}
\mbox{\includegraphics*[width=3.in, angle=0]{Fig4b.eps}}
\end{center}
\caption{Phase diagram of two binary mixtures of HEs  
with $(\kappa_1,\kappa_2)=(6,2)$ (a) and 
$(\kappa_1,\kappa_2)=(5,2)$ (b) while the particles areas of 
both species are equals. The meanings of different lines 
and labels coincide with those of Fig. \ref{fig1}. The tricritical and 
critical endpoints are  
shown with filled circle and square, respectively.
Finally, the critical point is shown with an 
open circle.} 
\label{fig2}
\end{figure}

The second study we have carried out concerns the calculation 
of the phase diagram of two 
similar binary mixtures of HEs, again with the same particle area 
$a_1=a_2$ but with different aspect 
ratios given by  $\kappa_1=6$ and $\kappa_2=2$ for the 
first mixture [Figure \ref{fig2} (a)] while they are 
$\kappa_1=5$ and $\kappa_2=2$ 
for the second mixture [Figure \ref{fig2} (b)]. There is an important
difference between 
both phase diagrams in that the first one 
has a region of I-N demixing above the tricritical point 
[see Figure \ref{fig2} (a)] 
which is replaced at higher pressures by a 
N-N demixing (just when the I-N spinodal intersects 
the I binodal of the I-N coexistence).
We call this point a critical end point. The I phase as usual 
is rich in HE species with small aspect ratio. The second phase diagram 
exhibits a second order I-N transition for all compositions and at 
higher pressures the system demix into two different N phases. 
This N-N demixing transition ends at a critical point. 

\begin{figure}
\begin{center}
\mbox{\includegraphics*[width=3.5in, angle=0]{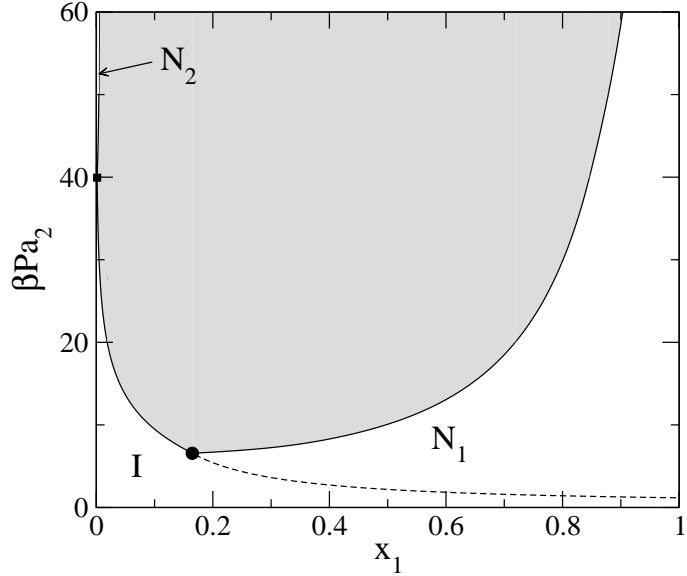}}
\end{center}
\caption{Phase diagram of a binary mixture of HEs 
with aspect ratios $\kappa_1=5$ and $\kappa_2=2$ while the 
partcle area $a_1=2.5a_2$. The meanings of labels 
and lines are as in Figigure \ref{fig1}.}
\label{fig3}
\end{figure}

How affects the difference in particle areas on the phase 
behavior constitutes the following point we have elucidated. 
The aspect ratios of both ellipses are again $\kappa_1=5$ 
and  $\kappa_2=2$ while the ratio between particle areas 
is $a_1/a_2=2.5$. 
In Fig. 
\ref{fig3} we show the resulting phase diagram. 
Again appears a window 
of I-N demixing but now the stability region of the 
second nematic (N$_2$) 
shrinks considerably.

\begin{figure}
\begin{center}
\mbox{\includegraphics*[width=3.5in, angle=0]{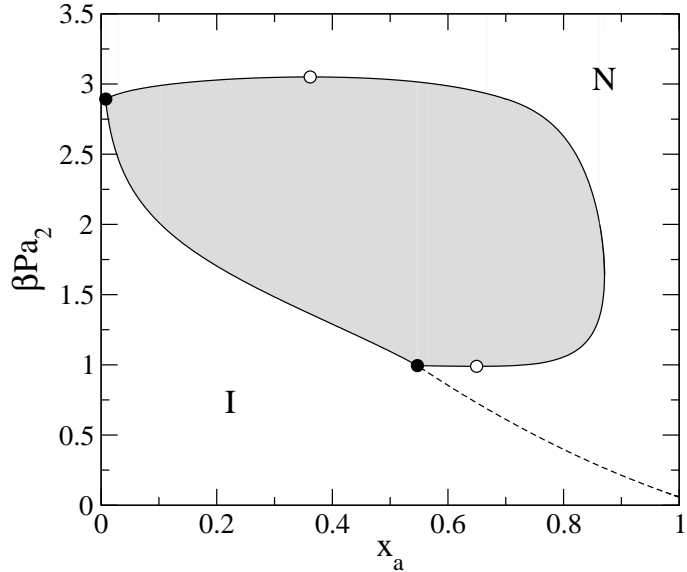}}
\end{center}
\caption{Mixture of HEs with the same aspect ratios 
$\kappa_1=\kappa_2=5$ and different particle areas 
$a_1=50a_2$. The critical and tricritical 
points are shown with open and filled circles respectively.} 
\label{fig4}
\end{figure}

Now we consider a mixture of HEs with the same aspect ratios 
$\kappa_1=\kappa_2=5$ and different areas, specifically $a_1/a_2=50$. 
The phase diagram is shown in Figure \ref{fig4}.  
This mixture exhibits two tricritical points defining 
the lower and upper limits of the I-N demixing and two 
critical points defining the proper limits of the N-N phase separation.  
In Figure \ref{fig4} the molar fraction of the first species is 
substituted by its area fraction, i.e. $x_a=x_1a_1/(x_1a_1+x_2a_2)$. 
In such a way the position of all these points can be properly discerned.

\subsection{Mixtures of HE and HR}
\label{cross}
\begin{figure}
\begin{center}
\mbox{\includegraphics*[width=3.5in, angle=0]{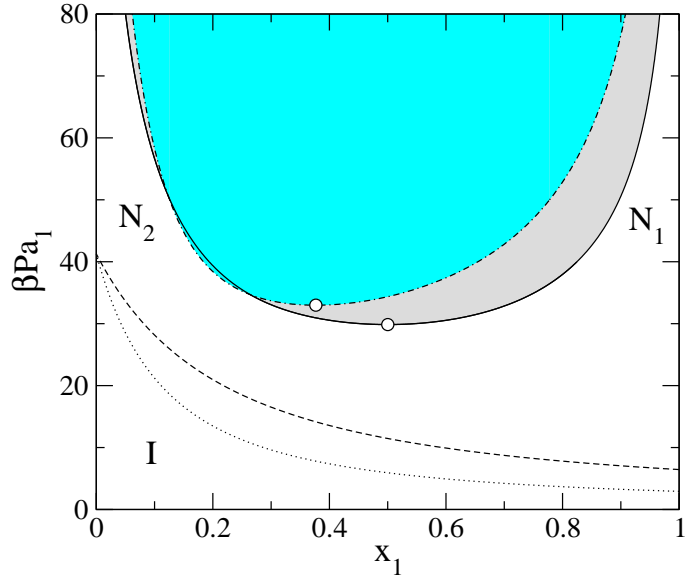}}
\end{center}
\caption{A mixture of HRs (species 1) and HEs (species 2) 
with the same particle areas $a_1=a_2$ while the aspect ratios 
are $\kappa_1=5$ and $\kappa_2=2$, respectively. The dashed line represents the 
I-N spinodal while the solid lines represent the N$_1$-N$_2$ demixing transition. 
Also plotted is  
the phase diagram of Figure \ref{fig2} (b) corresponding 
to the mixture of HEs 
with the same particle aspect ratios and areas. 
The dotted and dashed-dotted lines represent the spinodal 
and demixing curves,respectively, for this case.} 
\label{fig5}
\end{figure}

This section is devoted to the study of mixtures of particles with 
different shapes, specifically HE- and HR-shaped particles. 
The first crossed--mixture we 
have studied is composed of species 
with aspect ratios coinciding with those of the already studied HE mixture 
with phase diagram plotted in Figure \ref{fig2} (b). Specifically   
the HR has the larger aspect ratio 
($\kappa_1=5$) while the HE has $\kappa_2=2$, and the areas of booth 
particles coincide. The aspect ratio of the rectangular particles 
is selected so as to be large enough to forbid the T phase stability \cite{M-R2}.  
In Figure \ref{fig5} we show the 
phase diagram of this mixture. In the same figure  
we plot for a better comparison the phase 
diagram of Figure \ref{fig2} (b).  
We can extract two main differences between both phase diagrams. 
Firstly, the I-N transition occurs at lower pressures for the HE
mixture. This feature is related to the change in the 
excluded area as the particle aligns preferentially along the nematic 
director. This change is larger in mixtures of HEs compared 
with crossed--mixtures. The 
other main difference is related to the size of the N-N demixing gap.  
The crossed--mixture has the wider gap which can be understood in terms 
of packing of particles at high pressures: particles with the same shape 
pack better than those with different shapes. This in turn has already 
been quantified in Section \ref{NN} through the coefficients $q_{ee}$ and $q_{er}$ 
(see Figure \ref{fig0}).

\begin{figure}
\begin{center}
\mbox{\includegraphics*[width=3.5in, angle=0]{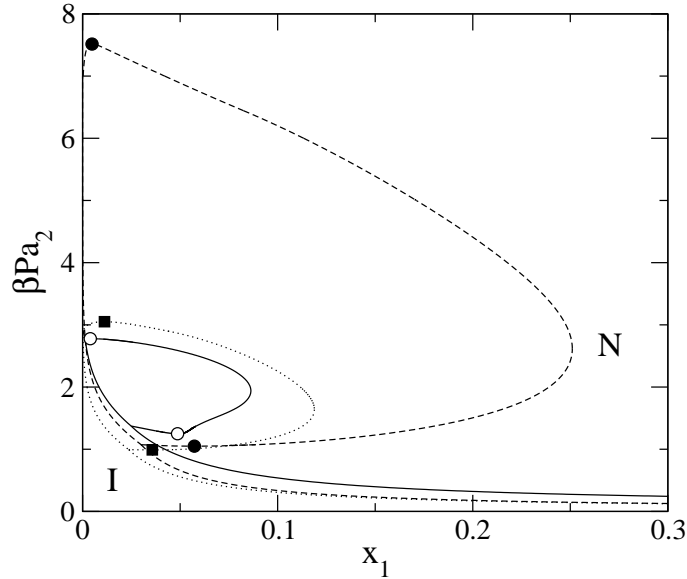}}
\end{center}
\caption{Phase diagrams of the binary mixtures with
equal particle aspect ratios  
($\kappa_1=\kappa_2=5$) but different  
areas ($a_1=50a_2$).  
The dotted lines represent the mixture of HEs, 
the solid lines represent the mixture of HRs (species 1) and 
HEs (species 2) and the dot-dashed lines represent the mixture of HEs (species 1) 
and HRs (species 2). The positions of critical points are shown with 
different symbols.}  
\label{fig6}
\end{figure}

The last study we have carried out is related to the effect that 
different particle shapes have on the  
phase diagrams of binary mixtures with species having very dissimilar  
areas (specifically $a_1=50a_2$) 
but equal aspect ratios (which were fixed to $5$). 
To this purpose we have calculated the phase diagrams of a mixture of 
HEs (the dotted lines in Figure \ref{fig6}), a crossed--mixture of 
HEis and HRs with rectangles being the large  species 
(the solid line in Figure \ref{fig6}) 
and the opposite mixture with
the ellipses being the large species (the dot-dashed line in Figure \ref{fig6}). 
As we can see from this figure the lower I-N spinodal corresponds to the 
mixture of HEs which is related to the major gain in 
the excluded area by the particle alignment of the most convex 
species, the HEs. For the same reason, we expect that the I-N 
demixing region is widened for crossed-mixtures with 
ellipses being 
the large species. The mixture of HEs and HRs, the later 
being the large species, 
has the smallest demixing gap which is related to the fact that 
a great amount of large rectangles enhances the stability of the 
isotropic phase with respect to the nematic, due to the disordered 
effect that cause the rectangular corners. 
Finally, the scenario with an intermediate demixing gap occurs 
for mixtures of HEs.    

\section{Conclusions}
\label{conclusions}   
One of the main purposes of the present work has been to encourage the 
statistical mechanical study of two-dimensional mixtures of hard anisotropic 
particles as a starting point to understand the phenomenology inherent 
in mixtures of granular particles. While three-dimensional 
mixtures have been systematically studied via density functional 
calculations and by MC simulations, there is a lack of  
results on two-dimensional mixtures, with scarce theoretical works and 
no simulation results. 

We have shown here that the phenomenology inherent to these mixtures 
are as rich as the three-dimensional mixtures, 
with all possible demixing scenarios 
where the demixed phases have isotropic or nematic symmetries. The 
absence of I-I demixing and the presence of second order I-N transition 
constitute the main differences with respect to the three-dimensional case. 

While mixtures of hard particles with rectangular and 
disco-rectangular geometries have been recently studied with 
density functional tools \cite{M-R2,Heras}, the present work 
constitutes the first attempt to incorporate  
the elliptical geometry into these studies. For a given 
aspect ratio the ellipse is the most convex particle we can imagine. 
As we have shown here, this geometrical property is translated to the 
the phase diagram topologies of mixtures with one or both species being 
ellipses. We trust that the same kind of results can be found in 
mixtures of ellipsoidal and cylindrical 
granular rods, the Basmati rice being an 
important paradigm of the former. 

We should take the conclusions about the stability of uniform phases that the 
SPT predicts with certain caution. Phase transitions to 
non-uniform phases, such as smectic or crystalline phases, 
can occur at pressures bellow the maximum values depicted in the 
phase diagrams plotted here. Only a full minimization of a density 
functional with respect to the density profile $\rho({\bf r},\phi)$ 
(depending also on the spatial variables) can clarify this problem. 
A density functional for mixtures of HEs based on the 
Parsons approach can in principle 
be implemented for the general density profile inhomogeneities. 
However this functional treats the translational degrees 
of freedoms in a crude approximation, wich is a serious drawback for two-dimensional 
systems where the many-body correlations are very important.   
   
\section*{Acknowledgments}
This work is part of the research project MOSAICO and we acknowledge support from grant 
FIS2010-22047-C05-C04 from 
the Ministerio de Ciencia y Tecnolog\'{\i}a, and grant MODELICO-CM 
from Comunidad Aut\'onoma de Madrid (Spain).

\appendix
\section{Calculation of the excluded area between two ellipses}
\label{A_HE}
\begin{figure}
\begin{center}
\mbox{\includegraphics*[width=3.5in, angle=0]{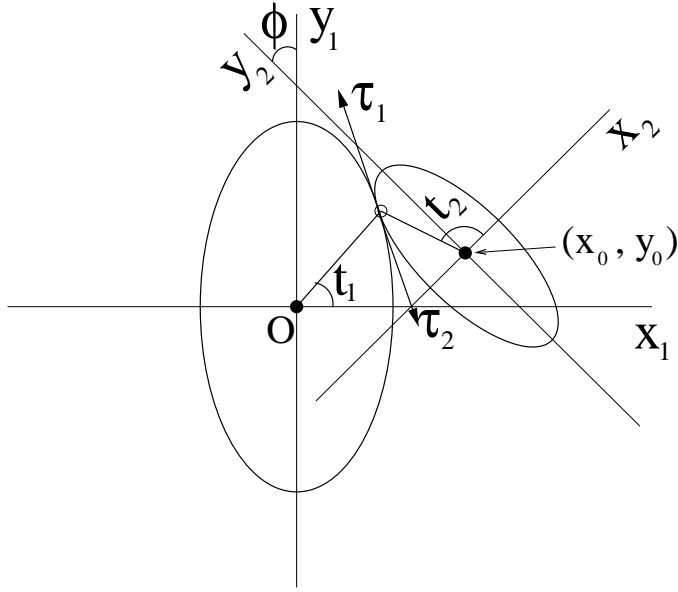}}
\end{center}
\caption{Configuration of the closest approach between two ellipses 
with semi-axes $\{{\sf a}_1,{\sf b}_1\}$ and $\{{\sf a}_2,{\sf b}_2\}$ with a relative   
orientation defined by the angle between their long axis, which is equal to $\phi$.}
\label{sketch1}
\end{figure}

In this section we calculate the excluded area between two different ellipses. 
The center of an  
ellipse with  semi-axis ${\sf a}_1$ and ${\sf b}_1$ will be placed at the origin 
of the coordinate system with the major semi-axis ${\sf b}_1$ being 
parallel to the 
$y$ Cartesian axis. The other ellipse, with semi-axes ${\sf a}_2,{\sf b}_2$, is 
placed at the point ${\bf r}_0=(x_0,y_0)$ and has its major axis 
forming and angle $\phi$ with respect to the long axis of the first 
ellipse. The vector ${\bf r}_0$ is such that there is only one common 
tangent point between both ellipses (see Figure \ref{sketch1}). 

The parametrised coordinates of the first ellipse are 
\begin{eqnarray}
{\bf r}_1(t_1)=[x_1(t_1),y_1(t_1)]=({\sf a}_1\cos t_1,
{\sf b}_1\sin t_1), \quad  t_1\in[0,2\pi],
\end{eqnarray}
while the second vector ${\bf r}_2(t_2)=[x_2(t_2),y_2(t_2)]$ 
is parametrically defined as   
\begin{eqnarray}
x_2(t_2)&=&x_0+{\sf a}_2\cos t_2\cos\phi-{\sf b}_2\sin t_2\sin\phi,\\
y_2(t_2)&=&y_0+{\sf b}_2\sin t_2\cos\phi+{\sf a}_2\cos t_2\sin\phi
\end{eqnarray}
with $t_2\in[0,2\pi]$. The tangent unit vectors $\boldsymbol{\tau}_i$ 
corresponding to both ellipses, calculated by differentiating 
$[x_i(t_i),y_i(t_i)]$ with respect to the parameters $t_i$, are 
\begin{eqnarray}
\hspace*{-0.5cm}
\boldsymbol{\tau}_1(t_1)=\frac{1}{\tau_1}(-{\sf a}_1\sin t_1,{\sf b}_1\cos t_1),\\
\hspace*{-0.5cm}
\boldsymbol{\tau}_2(t_2)=\frac{1}{\tau_2}(-{\sf a}_2\sin t_2\cos\phi-{\sf b}_2\cos t_2\sin\phi,
{\sf b}_2\cos t_2\cos\phi-{\sf a}_2\sin t_2\sin\phi),
\end{eqnarray}
where $\tau_i=\sqrt{{\sf a}_i^2\sin^2t_i+{\sf b}_i^2\cos^2t_i}$. The condition 
of a common tangent point between both ellipses can be expressed 
by the following set of equations: 
\begin{eqnarray}
{\bf r}_1(t_1)&=&{\bf r}_2(t_2),\\
\boldsymbol{\tau}_1(t_1)&=&-\boldsymbol{\tau}_2(t_2),
\end{eqnarray}
which can be solved to obtain $\sin t_2$ (or $\cos t_2$) as a function of 
$t_1$ and also for $x_0(t_1)$ and $y_0(t_1)$ with the result 
\begin{eqnarray}
\sin t_2&=&\frac{{\sf b}_2({\sf b}_1\cos t_1\sin\phi-{\sf a}_1\sin t_1\cos\phi)}
{\delta(t_1)^{1/2}},\\
\cos t_2&=&\frac{-{\sf a}_2({\sf a}_1\sin t_1\sin\phi+
{\sf b}_1\cos t_1\cos\phi)}
{\delta(t_1)^{1/2}},\\
x_0(t_1)&=&{\sf a}_1\cos t_1-{\sf a}_2\cos t_2\cos\phi+{\sf b}_2\sin t_2\sin\phi,\\
y_0(t_1)&=&{\sf b}_1\sin t_1-{\sf b}_2\sin t_2\cos\phi-{\sf a}_2\cos t_2\sin\phi,
\end{eqnarray}
where 
\begin{eqnarray}
\hspace*{-1.cm}
\delta(t_1)={\sf a}_2^2({\sf a}_1^2\sin^2t_1+{\sf b}_1^2\cos^2t_1)+
({\sf b}_2^2-{\sf a}_2^2)({\sf b}_1\cos t_1\sin\phi-{\sf a}_1
\sin t_1\cos\phi)^2 \nonumber \\
\hspace*{-0.2cm}
={\sf b}_2^2({\sf a}_1^2\sin^2t_1+{\sf b}_1^2\cos^2t_1)-
({\sf b}_2^2-{\sf a}_2^2)({\sf a}_1\sin t_1\sin\phi+
{\sf b}_1\cos t_1\cos\phi)^2,
\end{eqnarray}
The excluded area $A(\phi)$ can be calculated as  
\begin{eqnarray}
A(\phi)&=&\frac{1}{2}\int_0^{2\pi}\left[y_0'(t_1)x_0(t_1)-y_0(t_1)x_0'(t_1)
\right]dt_1,
\end{eqnarray}
After some long calculations we arrive at the following relatively 
simple formula for this function:
\begin{eqnarray}
A(\phi)=\pi({\sf a}_1{\sf b}_1+{\sf a}_2{\sf b}_2)+2\left[\Delta_+(\phi)+\Delta_-(\phi)\right]
E[\kappa(\phi)],
\label{final}
\end{eqnarray}
where $E(\kappa)=\int_0^{\pi/2}\sqrt{1-\kappa\sin^2\theta}d\theta$ is 
the complete elliptic integral of the second kind and where we have 
defined   
\begin{eqnarray}
\kappa(\phi)&=&\frac{4\Delta_+(\phi)\Delta_-(\phi)}
{\left[\Delta_+(\phi)+\Delta_-(\phi)\right]^2},\\
\Delta_{\pm}(\phi)&=&\sqrt{({\sf b}_1{\sf b}_2\pm {\sf a}_1{\sf a}_2)^2-
({\sf b}_2^2-{\sf a}_2^2)({\sf b}_1^2-{\sf a}_1^2)
\cos^2\phi}.
\end{eqnarray}

\section{Ellipse-rectangle excluded area}
\label{A_HE_HR}

\begin{figure}
\begin{center}
\mbox{\includegraphics*[width=2.2in, angle=0]{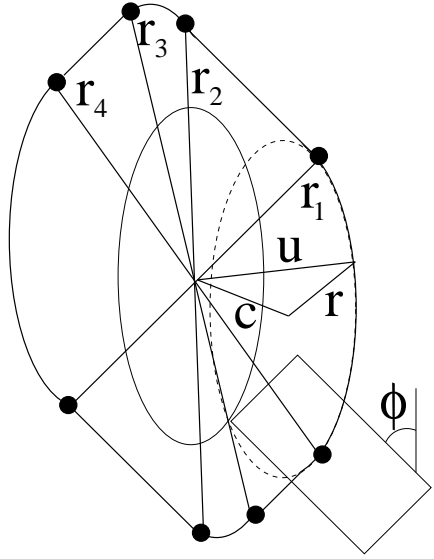}}
\end{center}
\caption{The excluded area between an ellipse with semi-axes ${\sf a}$ and 
${\sf b}$ and 
a rectangle of width $\sigma$ and length $L$. Different sections of the 
excluded area are shown.}
\label{fig2a}
\end{figure}

The purpose of this section is to obtain an expression for the excluded area 
between an ellipse with semi-axes ${\sf a}$ and ${\sf b}$ (${\sf b}>{\sf a}$) and a rectangle with 
length and width equal to $L$ and $\sigma$, respectively. We show in 
Figure \ref{fig2} an schematic representation of the excluded area. 
It can be seen from the figure that the total area can be computed by 
summing two different contributions: (i) one coming from four triangles (two 
of them having different areas $A_i^{(t)}$, $i=1,2$) and (ii) the 
second, coming from four figures with elliptical borders (two of them having 
areas $A_i^{(e)}$, $i=1,2$). The areas corresponding to the triangles 
are 
\begin{eqnarray}
A_1^{(t)}&=&\frac{1}{4}\sqrt{4(r_1r_2)^2-(r_1^2+r_2^2-L^2)^2},\\
A_2^{(t)}&=&\frac{1}{4}\sqrt{4(r_3r_4)^2-(r_3^2+r_4^2-\sigma^2)^2},
\end{eqnarray}
where $r_i$ are the absolute values of the vectors shown in Figure \ref{fig2}.
These areas can be computed as 
\begin{eqnarray}
A_1^{(t)}&=&\frac{L}{2}\left[\frac{\sigma}{2}+\sqrt{{\sf a}^2\cos^2\phi+
{\sf b}^2\sin^2\phi}
\right],\\ 
A_2^{(t)}&=&\frac{\sigma}{2}\left[\frac{L}{2}+\sqrt{{\sf a}^2\sin^2\phi+
{\sf b}^2\cos^2\phi}
\right], 
\end{eqnarray}
 
To compute the areas of figures with the 
elliptical borders we define the radius vector ${\bf u}$ 
from the origin to the border of the excluded area 
(see Figure \ref{fig2a}), which in turns is equal 
to the sum of two vectors: the vector ${\bf c}=(c_x,c_y)$ which accounts 
for the position of the center of mass of the rectangle with its inner vertex 
touching the perimeter of the ellipse and the vector ${\bf r}=
({\sf a}\cos\theta,{\sf b}\sin\theta)$ which has been parametrised with the angle 
$\theta\in[\theta_1,\theta_2]$. Thus, the area of this figure 
can be calculated from 
\begin{eqnarray}
A^{(e)}=\frac{1}{2}\int_{\psi_1}^{\psi_2}u^2(\psi)d\psi=
\frac{1}{2}\int_{\theta_1}^{\theta_2}u^2(\theta)\psi'(\theta)d\theta,
\label{pp}
\end{eqnarray}
with $\psi$ the angle that the vector ${\bf u}$ forms with the $x$-axis. 
Taking into account the fact that $u=\sqrt{({\sf a}\cos\theta+c_x)^2+
({\sf b}\sin\theta+c_y)^2}$ and 
\begin{eqnarray}
\sin\psi=u^{-1}\left(c_y+{\sf b}\sin\theta\right),\quad 
\cos\psi=u^{-1}\left(c_x+{\sf a}\cos\theta\right),
\end{eqnarray}
we find from (\ref{pp}) that
\begin{eqnarray}
\hspace*{-0.5cm}
A^{(e)}_i=\frac{1}{2}\left[{\sf a}{\sf b}\left(\theta_2^{(i)}-\theta_1^{(i)}\right)+
{\sf b}c_x\left(\sin\theta_2^{(i)}-\sin\theta_1^{(i)}\right)
-{\sf a}c_y^{(i)}\left(\cos\theta_2^{(i)}-\cos\theta_1^{(i)}\right)\right],
\end{eqnarray}
where $c_{x,y}^{(i)}$ are the components 
of the vector ${\bf c}^{(i)}$ describing the movement of the center 
of mass of the rectangle inside the two different regions with elliptical 
borders, while $\theta_{1,2}^{(i)}$ are the lower and upper limits of 
the parameter $\theta$ inside these regions. All these quantities 
are explicit functions of ${\sf a},{\sf b}$, $\sigma$ and $L$. 
It can be easily shown that 
\begin{eqnarray}
\hspace*{-1.cm}
\sum_i A^{(e)}_i=\frac{\pi}{2}{\sf a}{\sf b}+\frac{L}{2}\sqrt{{\sf a}^2\cos^2\phi+
{\sf b}^2\sin^2\phi}+\frac{\sigma}{2}\sqrt{{\sf a}^2\sin^2\phi+{\sf b}^2\cos^2\phi}. 
\end{eqnarray}
Thus, the total excluded area can be obtained by summing all the 
contributions, 
i.e. $A(\phi)=2\sum_i\left[A_i^{(t)}+A_i^{(e)}\right]$, with the final result
\begin{eqnarray}
\hspace*{-1.0cm}
A(\phi)=L\sigma+\pi {\sf a}{\sf b}+2\left[
\sigma\sqrt{{\sf a}^2\sin^2\phi+{\sf b}^2\cos^2\phi}+L
\sqrt{{\sf a}^2\cos^2\phi+
{\sf b}^2\sin^2\phi}\right].
\end{eqnarray}

\section{Minimization of the Gibbs free-energy per particle}
\label{minimiza}
As we have already pointed out before, 
the double angular average  
$\langle\langle A^{(0)}_{\mu\nu}\rangle\rangle$ is the main 
ingredient of SPT which can be computed as  
\begin{eqnarray}
\langle\langle A_{\mu\nu}^{(0)}\rangle\rangle 
=\int_0^{\pi}d\phi\Psi_{\mu\nu}(\phi)A_{\mu\nu}^{(0)}(\phi),
\end{eqnarray}
where the functions 
\begin{eqnarray}
\Psi_{\mu\nu}(\phi)=\int_0^{\pi}d\phi' h_{\mu}(\phi')h_{\nu}(\phi+\phi'),
\end{eqnarray}
were defined, which using Equation (\ref{ins}) become 
\begin{eqnarray}
\Psi_{\mu\nu}(\phi)&=&\frac{I_0[\xi_{\mu\nu}(\phi)]}
{\pi I_0(\lambda_{\mu})I_0(\lambda_{\nu})},\\
\xi_{\mu\nu}(\phi)&=&\sqrt{\lambda_{\mu}^2+\lambda_{\nu}^2+
2\lambda_{\mu}\lambda_{\nu}\cos 2\phi}.
\end{eqnarray}
To numerically implement the minimization of the Gibbs free energy per 
particle $g$ with respect to the parameters $\lambda_{\mu}$ we 
have used the following expressions for the derivative of $g$ with 
respect to $\lambda_{\mu}$
\begin{eqnarray}
\frac{\partial g}{\partial\lambda_{\mu}}&=&
x_{\mu}\left\{\lambda_{\mu}\left[1-\frac{I_1^2(\lambda_{\mu})}
{I_0^2(\lambda_{\mu})}\right]-\frac{I_1(\lambda_{\mu})}
{I_0(\lambda_{\mu})}+y\sum_{\nu}x_{\nu}S_{\mu\nu}\right\},\label{uu}\\
S_{\mu\nu}&=&\int_0^{\pi}d\phi\frac{\partial \Psi_{\mu\nu}(\phi)}
{\partial\lambda_{\mu}}A^{(0)}_{\mu\nu}(\phi),
\end{eqnarray}
where   
\begin{eqnarray}
\hspace*{-0.5cm}
\frac{\partial\Psi_{\mu\nu}(\phi)}{\partial\lambda_{\mu}}
=\frac{1}{\pi I_0(\lambda_{\nu})I_0^2(\lambda_{\mu})}
\left[(\lambda_{\mu}+\lambda_{\nu}\cos 2\phi)\frac{I_1(\xi_{\mu\nu})}
{\xi_{\mu\nu}}I_0(\lambda_{\mu})-I_0(\xi_{\mu\nu})I_1(\lambda_{\mu})
\right].
\end{eqnarray}
We have solved the  
set of two non-linear equations $\partial g/\partial \lambda_{\mu}=0$ 
($\mu=1,2$) with respect to 
two unknowns $(\lambda_1,\lambda_2)$.

\end{document}